\documentclass[aps, 10pt,prb,superscriptaddress,twocolumn, floatfix]{revtex4-1}
\usepackage{amssymb}
\usepackage{amsmath}
\usepackage{bbold}
\usepackage{braket}
\usepackage{silence}
\DeclareMathOperator{\Tr}{Tr}
\usepackage{color}
\usepackage{graphicx}

\usepackage{hyperref}
\usepackage{lipsum}
\usepackage{float}
\usepackage{mathtools}

\begin{document}
\title{Robust zero-energy bound states in a helical lattice}
\author{Pengke Li}
\affiliation{Department of Physics and Center for Nanophysics and Advanced Materials, U. Maryland, College Park, MD}
\author{Jay D. Sau}
\affiliation{Condensed Matter Theory Center, Joint Quantum Institute and Department of Physics, U. Maryland, College Park, MD}
\author{Ian Appelbaum}
\affiliation{Department of Physics and Center for Nanophysics and Advanced Materials, U. Maryland, College Park, MD}

\begin{abstract}
Atomic-scale helices exist as motifs for several material lattices. We examine a tight-binding model for a single one-dimensional monatomic chain with a $p$-orbital basis coiled into a helix. A topologically nontrivial phase emerging from this model supports a zero-energy mode localized to a boundary, always embedded within a continuum band, regardless of termination site. We identify a topological invariant for this phase that is related to the number of zero energy end modes by means of the bulk-boundary correspondence, and give strict conditions for the existence of the bound state. Another, non-topological, gapped edge mode in the model spectrum has practical consequences for surface states in e.g. trigonal tellurium and selenium and other van~der~Waals-bonded one-dimensional semiconductors.
\end{abstract}
\maketitle

\section{Introduction}
The helix is unique among geometric curves, due to the inequivalent chirality of left-handed and right-handed mirror images. Helical structure is fairly common among organic molecules, such as DNA, RNA and a variety of proteins. Interestingly, in many biological processes, one enantiomer dominates over the other due to symmetry breaking and autoamplification. Helical atomic order is also seen in some inorganic solids, such as in the well-known case of optically active quartz. 

Whereas the chiral quartz lattice is formed from achiral building blocks (Si-O tetrahedra), an alternative class of optically-active materials is formed from van~der~Waals-bonded, aligned atomic helix chains that are themselves chiral. This class of materials includes the transition metal binary $\alpha$-HgS ``cinnabar'' \cite{Glazer_JAC86}, and trigonal crystals of the elemental chalcogenide semiconductors selenium (Se) and tellurium (Te). The chirality of the latter two distinguishes them from the vast majority of other pure elemental crystals that possess inversion symmetry.

Regardless of specific geometry, the lattice structure determines the nature of electronic states residing within. In addition to the usual delocalized Bloch waves, finite lattices can support surface states localized to a boundary and that decay into the bulk, a topic pioneered by Tamm and Shockley in the 1930s.\cite{Tamm_PZSU32,Shockley_PR39} Later, the seminal work of Su, Schrieffer, and Heeger (SSH)\cite{Su_PRL79, Su_PRB80, Heeger_RMP88} provided a successful model for dimerized polyacetylene, and introduced the concept of `topological' excitations, extending to bound edge modes of finite chains. However, despite the aforementioned prevalence of helical order in many natural materials systems, edge states of the 1-dimensional helix have, to the best of our knowledge, escaped scrutiny.

In this paper, we use the nearest-neighbor tight binding approximation to study the edge states of a one-dimensional helix arising from the interplay of multiple $p$-orbitals, as appropriate for the valence electronic structure of a single atomic chain of elemental metalloids (e.g. Se or Te). We show that the electronic structure of such a helix is constrained by symmetry to be equivalent to that of a translationally-invariant model with one atom per unit cell and broken mirror symmetry. We focus on a zero-energy mode that resembles the `topological' midgap edge state in the SSH model, and use the transfer matrix technique to determine parameters that guarantee its existence. However, in contrast to dimerization which plays an essential role in the SSH model, all neighboring bonds in the helical chain are equivalent so the edge state is robust against edge termination phase. We find that our helix model is an example from a class of gapless topological phases that support zero-energy end modes protected from hybridization with the continuum. Its bulk invariant (with values in $\mathbf Z$) determines the number of end modes and is given by the number of zero-energy band crossings.

Using the Green's function technique, we examine the evolution of this edge state phase. With this numerical method, we additionally identify a pair of ordinary non-topological Tamm/Shockley states that only appear in a forbidden gap, indicating further relevance of our model to the behavior of many real helical materials.

\section{A Motivating Illustration}

\begin{figure}
\centering 
\includegraphics[scale=0.65]{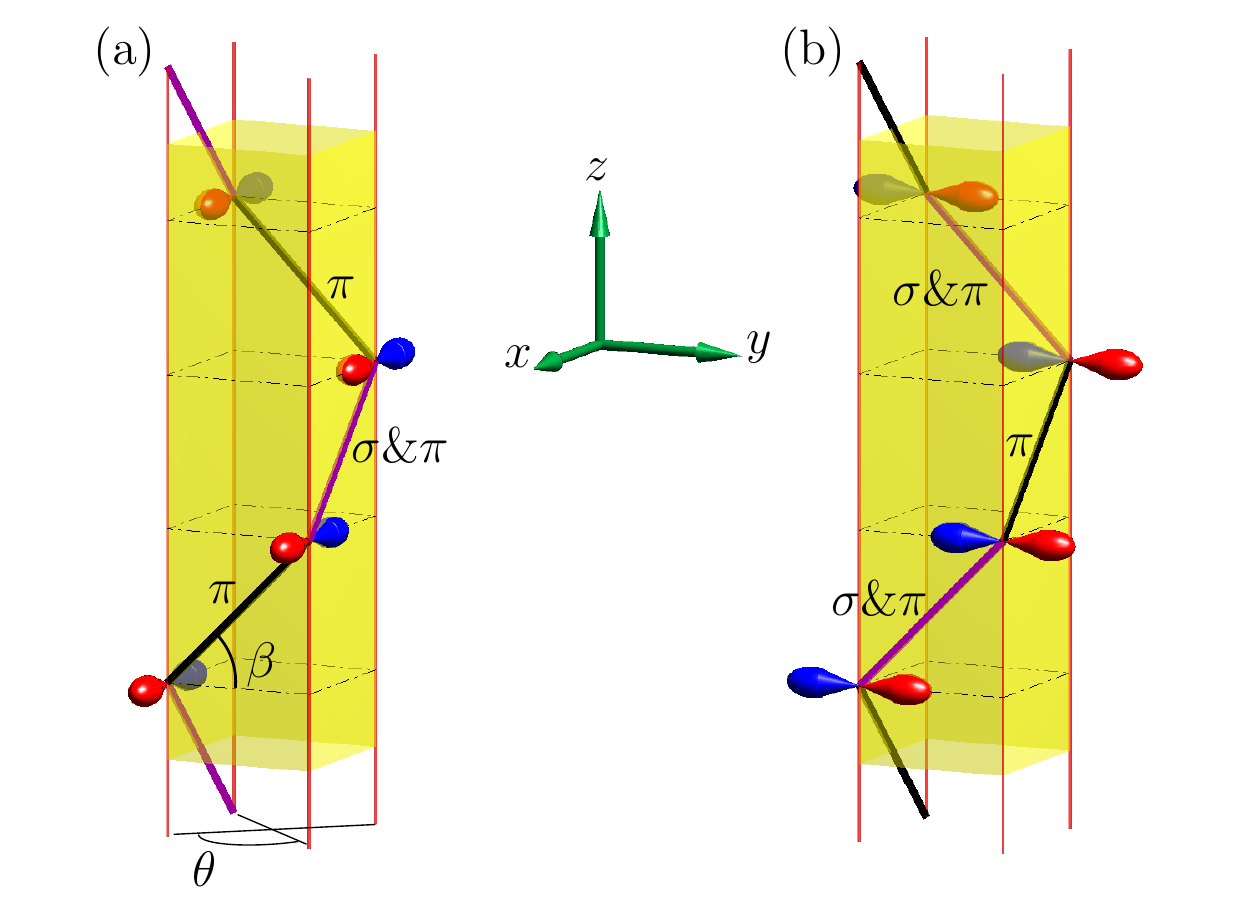}
\caption{Fourfold helical chains along the $z$-direction, with $p_x$-orbital (a) and $p_y$-orbital (b) on each site. Since $\theta = 90^\circ$, the nearest neighbor coupling of these two types of orbitals vanishes. Single unit cells are highlighted in yellow. The polarities of the orbitals are indicated by red and blue. The nearest neighbor covalent bonds are denoted by purple and black lines alternatively, corresponding to $\sigma-\pi$ mixed coupling and pure $\pi$ coupling, respectively. Note that the bond ordering is out-of-phase between the $p_x$ and $p_y$ orbitals. 
\label{fig:fourfold}}
\end{figure}

A single unit cell of a monatomic helical chain with fourfold rotational symmetry is shown in Fig. \ref{fig:fourfold}. The choice of fourfold symmetry simplifies the tight-binding description: If each lattice site hosts only $p_x$ and $p_y$ orbitals, these basis states are uncoupled, and bonding between nearest-neighbor sites alternates between pure $\pi$-like and mixed $\pi$- and $\sigma$-like. Importantly, wherever $p_x-p_x$ bonding is $\pi$-like [black lines in Fig.~\ref{fig:fourfold}(a)], $p_y-p_y$ bonding between the same lattice sites is a superposition of $\pi$- and $\sigma$-like [purple lines in Fig.~\ref{fig:fourfold}(b)], and vice-versa.

Because two different orbitals on adjacent lattice sites remain orthogonal, the tight-binding Hamiltonian for this system can be trivially block-diagonalized into two entirely decoupled chains: one with bonds only between $p_x$ orbitals and one with only $p_y$ orbitals.  Crucially, these two independent chains are nevertheless {\em{correlated}} out of phase, so that if one terminates with a pure $\pi-$like bond, the other terminates with a mixed $\pi-$ and $\sigma$-like bond.

As a result, this two-band model is equivalent to the spatial superposition of two SSH dimerized chains, where one or the other is guaranteed to satisfy the condition for a ``topological" mid-gap zero-energy mode,  {\em{regardless}} of specific termination location. This robust behavior is distinct from the single SSH chain, which must terminate with the weaker of the two bond types to have a topological state bound there.

Despite its enticingly indelible topological bound state, this fourfold helix model is clearly unphysical with only two orbitals, although approximate fourfold helices do indeed exist in solids\cite{Hirschle_ACC01}. In addition to inclusion of alternative rotational symmetry, a physically-relevant model should at least incorporate the otherwise-degenerate $p_z$ orbital state at each atomic lattice site (if not the energetically-remote $s$ and $d$ orbitals). However, doing so will induce second-order off-diagonal coupling between the $p_x$ and $p_y$ orbitals, making decomposition into two decoupled but correlated chains (as we describe above) formally impossible. 

The following questions naturally arise. Do the topological features of the conceptually simple but unphysical $p_x-p_y$ fourfold helical chain model persist even after a more complete basis is considered? How sensitive are they to the geometric parameters of the helix, i.e. specific rotational symmetry and winding pitch (determined by angles $\theta$ and $\beta$ shown in Fig. \ref{fig:fourfold}, respectively)? Do these topological properties play a relevant role in physical systems such as the electronic structure of trigonal phases of covalently-bonded elemental semiconductors (such as selenium and tellurium\cite{Doi_JPSP70, Joannopoulos_PRB75,Hirayama_PRL15, Yoda_SciRep15})? 

The remainder of this paper is dedicated to addressing these questions. In the next section, Sec. \ref{sec:model}, we derive a tight-binding hopping matrix for a general  helix of $p$-orbitals that is subsequently used to construct appropriate Hamiltonians. We analyze the properties of this Hamiltonian and its hopping matrix in Sec. \ref{sec:zero}, determining the conditions for existence of the zero-energy bound state. In Sec. \ref{sec:num} we present numerically-computed example spectra and explore their features in light of the analysis in the preceding. We discuss our conclusions and suggest implications for organic polymers in Sec. \ref{sec:conc}. 

\section{Model Hamiltonian\label{sec:model}}
To capture the generic features regardless of the specific geometric parameters, we construct the tight-binding coupling matrix by considering the helix as a strictly 1-d wire with a position dependent, {\em{and}} rotated basis. As shown below, this approach reduces the unit cell to a single site, even for irrational values of the rotation which would otherwise yield indefinitely-large standard unit cells.

For a straight wire along the $z$-axis, this coupling matrix $t=D$, where $D$ is exactly diagonal in the $\{p_x,p_y,p_z\}$ basis, and the diagonal elements are the usual tight-binding parameters $-V_{\pi},-V_{\pi},V_{\sigma}$. Here, the signs are chosen so that bonding is favored by large (positive) values of the tight-binding parameters.

When the chain is coiled into a helix, hopping is both a rotation with operator $R_z(\theta)$ times the bare coupling $q=\hat r \otimes \hat r (V_{\sigma}+V_{\pi})-1\cdot V_{\pi}$, where $\hat r$ is the nearest-neighbor unit vector and we have exploited the fact that $\{p_x,p_y,p_z\}$ transform as the components of a polar vector. Choosing $\hat{r}=
\begin{bmatrix}
0, \cos\beta, \sin\beta
\end{bmatrix}^\dagger,$ 
\begin{align} 
q=
\begin{bmatrix}
0&0&0\\
0&\cos^2\beta&\sin\beta\cos\beta\\
0&\cos\beta\sin\beta&\sin^2\beta
\end{bmatrix}(V_{\sigma}+V_{\pi})-1\cdot V_{\pi},\notag
\end{align}
where $\beta$ is the angle between $\hat r$ and the $x-y$ plane. Notice that if $\beta=\pi/2$, we recover the trivial coupling in a linear chain along $z$, and $q=D$ as expected. Since the only off-diagonal matrix elements are between $y$ and $z$, $q$ can be diagonalized by a unitary transformation with the rotation $R_x(\pi/2-\beta)=\tilde{R}_x(\beta)$. We thus have coupling matrix elements
\begin{align}
 t_{n,n+1}=&\langle\phi_n|R_z(\theta)\tilde{R}_x(\beta)D\tilde{R}_x(\beta)^\dagger|\phi_{n+1} \rangle\notag\\
 =&
\langle \psi_n|\tilde{R}_x(\beta)^\dagger R_z(\theta)\tilde{R}_x(\beta)D|\psi_{n+1} \rangle,\notag
\end{align}
where we define a new basis $\ket{\psi}=\tilde{R}_x(\beta)^\dagger \ket{\phi}$. This yields a coupling matrix $t= \tilde{R}_x(\beta)^\dagger R_z(\theta)\tilde{R}_x(\beta)D=$
\begin{align}
\resizebox{0.9\hsize}{!}{
$ \begin{bmatrix}
-V_{\pi}\cos\theta & V_{\pi}\sin\beta\sin{\theta} &    V_{\sigma}\cos\beta\sin{\theta}\\
-V_{\pi}\sin\beta\sin{\theta} &     V_{\pi}(2\sin^2\frac{\theta}{2}\sin^2\beta-1)& V_{\sigma}\sin 2\beta\sin^2\frac{\theta}{2}\\
 V_{\pi}\cos\beta\sin{\theta} & -V_{\pi}\sin 2\beta\sin^2\frac{\theta}{2} &V_{\sigma}(1-2\sin^2\frac{\theta}{2}\cos^2\beta)
 \end{bmatrix},$}
\label{eq:t}
\end{align}
which has the obvious advantage that only one tight-binding parameter determines each matrix element, due to the fact that $D$ is diagonal and only multiplied on the left. Note that inversion asymmetry is captured by terms odd in $\beta$.

The above transformation corresponds to a basis in which one $p$-orbital is aligned to a bond in the forward-hopping direction (third row and column), another lies in the direction perpendicular to the plane formed by that bond and the helix radius (first row and column), and the third is orthogonal to both of the other two orbitals (second row and column).

\section{Zero-energy bound state \label{sec:zero}}

In a perfect helix without disorder and spin-orbit interaction, on-site energy vanishes, and the nearest-neighbor tight-binding Hamiltonian only involves hopping. The corresponding Schr\"odinger equation is then
\begin{align}
t\psi_{n+1}+t^\dagger\psi_{n-1}=\varepsilon \psi_n,
\label{eq:schr}
\end{align}
where $n=1,2,\dots$ is lattice site index, $\psi$ is a 3-element column vector, and $\varepsilon$ is the energy eigenvalue. Investigation of this equation can yield analytic information on the possible existence and properties of any bound state. 

\subsection{Transfer matrix at zero energy\label{sec:transfer}}
Of particular interest is a zero-energy bound state, similar to the SSH model. At $\varepsilon=0$ and $n=1$ (the terminal site), Eq. \ref{eq:schr} reads $t\psi^{\varepsilon=0}_2=0$, and for all other $n$ it yields $\psi^{\varepsilon=0}_{n+1}=M\psi^{\varepsilon=0}_{n-1}$, where the transfer matrix 
\begin{align}
M=-t^{-1}t^\dagger
\label{eq:M}
\end{align}
is nonsingular since $\det t=V_{\pi}^2V_{\sigma}\neq 0$, and we have $\psi^{\varepsilon=0}_n=0$ for {\em{all}} even $n$.  This bipartite sublattice character is a manifestation of the chiral symmetry\cite{Ryu_NJP10} (i.e., absence of even-order hopping terms) of our helical chain model, evident from the absence of on-site terms in the Hamiltonian operator, Eq. \ref{eq:schr}; for every state with energy $\varepsilon$, there is another at energy $-\varepsilon$ with opposite phase at every second site.

The analysis above does not guarantee that all states at zero energy are bound and localized to the chain endpoint. However, details of the transfer matrix given in Eq. \ref{eq:M} can reveal the conditions for its existence. First, note that $\det(M)=-1$ and its eigenvalues are $\{-1,z,z^{-1}\}$ (See Appendix for rigorous proof). Either $|z|=1$ and there is no localized bound state at zero energy, or else $\min\{|z|,|z^{-1}|\}<1$ controls the decay of the wavefunction into the bulk. To distinguish between these two cases, we can alternatively demand the condition $\{\Tr(M)<-3\}\oplus\{ \Tr(M)>1\}$ for the presence of the zero-energy state. The converse statement is more succinct: The criteria for the absence of bound state at zero energy, determined by brute calculation of $\Tr(M)$, is 
\begin{align}
0<F(\theta,\beta,\eta)<1,\label{eq:zerocondition}
\end{align}
where 
\begin{align}
&F(\theta,\beta,\eta)=\notag\\
&\frac{(1+\eta)^2}{4\eta}(\sin^2\beta\cos\theta+\cos^2\beta)^2-\frac{(1-\eta)^2}{4\eta}\cos^2\theta,\label{eq:F}
\end{align}
and $\eta=\frac{V_\sigma}{V_\pi}$. Only the ratio of tight-binding coefficients is important, reducing four free parameters to only three. 

\subsection{The function \texorpdfstring{$F(\theta,\beta,\eta)$}{F}}

\begin{figure}
\includegraphics{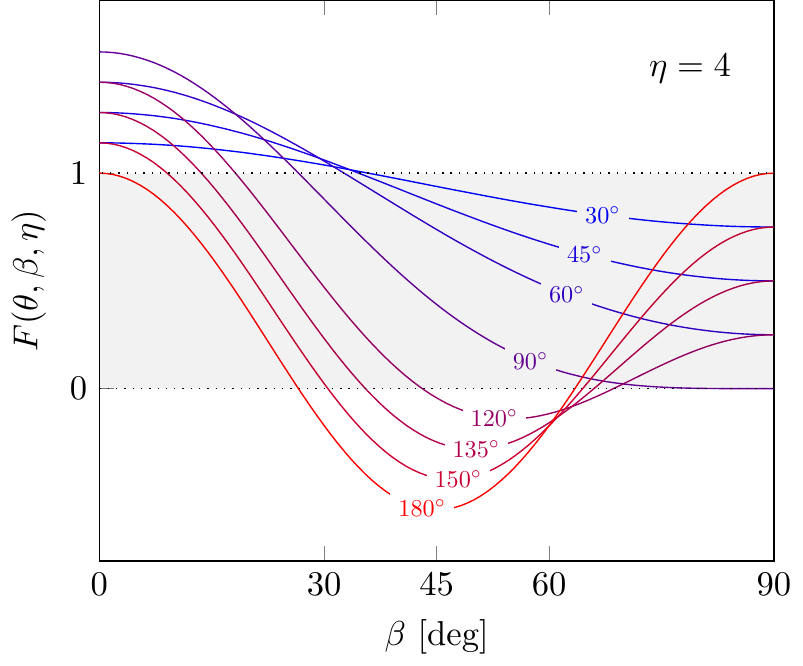}
\caption{Plots of Eq. \ref{eq:F} at fixed $\eta=4$, for several values of $\theta$. Shaded region indicates helical chains with no bound state at zero energy. \label{fig:F}}
\end{figure}

To understand the regimes where a zero-energy bound state is guaranteed, we plot Eq. \ref{eq:F} as a function of $\beta$, at fixed $\eta=4$, for several values of $\theta$ in Fig. \ref{fig:F}. Eq. \ref{eq:zerocondition} is violated for small values of $\beta$ so that $F>1$ and a zero-energy bound state exists, regardless of $0<\theta<180^\circ$. In the limit of small $\theta$, a lowest-order expansion of Eq. \ref{eq:F}  shows that the zero-energy bound state is present when $\beta<\arcsin\frac{\eta-1}{\eta+1}$; for $\eta=4$, this occurs at $\beta\approx 37^\circ$.

The bias toward small $\beta$ can be understood by investigating the matrix elements in Eq. \ref{eq:t} that couple the middle orbital to the other two that, when isolated, are responsible for an SSH-like mode. Specifically, all such matrix elements are proportional to $\sin\beta$, so that when $\beta$ is small, this coupling cannot destroy the robust zero-energy bound mode. 

For helices with $\theta>90^\circ$, another possible scenario for appearance of the zero-energy bound state occurs. A finite range of $\beta$ values violates Eq. \ref{eq:zerocondition} where $F<0$. As $\theta$ approaches $180^\circ$ (the pathological case of a two-fold helix where all sites lie in a plane), this region develops symmetrically around $\beta=45^\circ$. We discuss this special case in Sec. \ref{sec:conc}.

\subsection{Bulk-boundary correspondence}

It is clear from the previous section that the zero-energy state exists over a range of parameter-space that defines a phase. In the present section, we rigorously establish that the bulk bandstructure determines a topological invariant for this phase, which simultaneously allows us to calculate the number of these zero-energy modes.

We start with the general Bloch wavefunction $u\exp(ikna)\equiv u\lambda^n$ on a discrete lattice,
where the $N\times 1$ vector $u$ satisfies the Bloch equation, 
$[\lambda t+\lambda^{-1}t^\dagger]u=\varepsilon u$, 
which can be written as
\begin{align}
[\lambda^2 t+t^\dagger-\varepsilon\lambda] u=0.
\label{eq:Bloch}
\end{align}
Its characteristic equation clearly has $2N$ solutions for $\lambda$, giving wavefunctions of the general form 
\begin{align}
\psi_n=\sum_{1\leq m\leq 2N} c_m u_m \lambda_m^n.
\label{eq:psi}
\end{align}

The bound states we seek must be normalizable, so the sum in Eq. \ref{eq:psi} is restricted to $m$ for which $|\lambda_m|<1$, or equivalently $\Im(k_m)>0$. This eliminates all $2\ell$ unique real-valued $-\pi/a<k_m<+\pi/a$ which correspond to band crossings at $\varepsilon$ within the full Brillouin zone, leaving $2N-2\ell$ complex solutions. By considering the Hermitian conjugate of Eq. \ref{eq:Bloch}, we see that for every solution $k$ there is also a $k^*$, so that half of all these complex solutions are unphysical. 

The bound state Eq. \ref{eq:psi} is also subject to the boundary condition $\psi_0=0$, which is necessary to satisfy the Schr\"odinger equation (Eq. \ref{eq:schr}) at the chain end. Note that (only) when $\varepsilon=0$, solutions to Eq. \ref{eq:Bloch} come in pairs, so that for every $\lambda$ with associated eigenvector $u$, there is also a $-\lambda$ with the {\em{same}} $u$. A linear combination [Eq.~\ref{eq:psi}] that satisfies the boundary condition is $[(+\lambda)^n-(-\lambda)^n]u$. This odd superposition is identical to the zero-energy bipartite state obtained directly from the tight-binding equations. All  even superpositions are excluded, so that we have ultimately reduced the number of allowed bound states to at least $(N-\ell)/2$. As an example, for the case of 3 orbitals ($N=3$) in Eq.~(\ref{eq:t}), this gives one bound state per chain end when the corresponding infinite lattice has two band crossings in the full BZ ($\ell=1$). 

It is worth noting that when $N$ is odd, the spectrum of a chiral and translationally invariant  Hamiltonian cannot have a bulk gap at zero energy. Then, $\ell\neq 0$ at $\varepsilon=0$, making the bound state an instance of an ``embedded eigenstate" \cite{vonNeumann_PZ29,Hsu_NRM16}. It is thus an example of a one dimensional gapless topological state, so the usual topological invariant that describes chiral symmetric models such as SSH does not apply. Despite this, the model {\em{is}} topological in the sense that it is robust against perturbations as long as chiral symmetry and translational invariance are preserved. It also obeys a bulk-boundary correspondence, in the sense that the zero-energy state depends only on the number of orbitals and number of zero crossings of the bulk Hamiltonian. We can thus identify bulk band crossings at zero energy as the topological invariant of this phase. This quantity (as well as the number of end zero modes) can take any integer value and thus is a $\mathbf{Z}$ invariant.

Strict conditions determining the range of tight-binding parameters demanding a zero-energy mode depend on the details of the Bloch Hamiltonian and constituent matrix elements of $t$. Allowed values of $k_m$ are given by solution to $\det(\mathcal{H}(k)-\varepsilon)=0$,
where $\mathcal{H}(k)=\exp(ika) t+\text{h.c.}$ is the Bloch Hamiltonian. Thus at zero energy, for our $p$-orbital model and $3\times 3$ matrix $t$ given by Eq.~\ref{eq:t}, the characteristic equation is cubic in $\cos ka$, and given by 
\begin{align}
\cos ka\left[\cos^2 ka -F(\theta,\beta,\eta) \right]=0.\label{eq:blochreq}
\end{align}
One straightforward solution to Eq. \ref{eq:blochreq} is $\cos ka=0$, or $k=\pm\frac{\pi}{2a}$, corresponding to the midpoint from the Brillouin zone center to the zone edge. The conditions for real $k$ in the other two solutions (three band crossings at zero energy and no bound state) reduce once again to Eq. \ref{eq:zerocondition}.

\section{Numerical Spectra\label{sec:num}}
\begin{figure}
\includegraphics[scale=0.4]{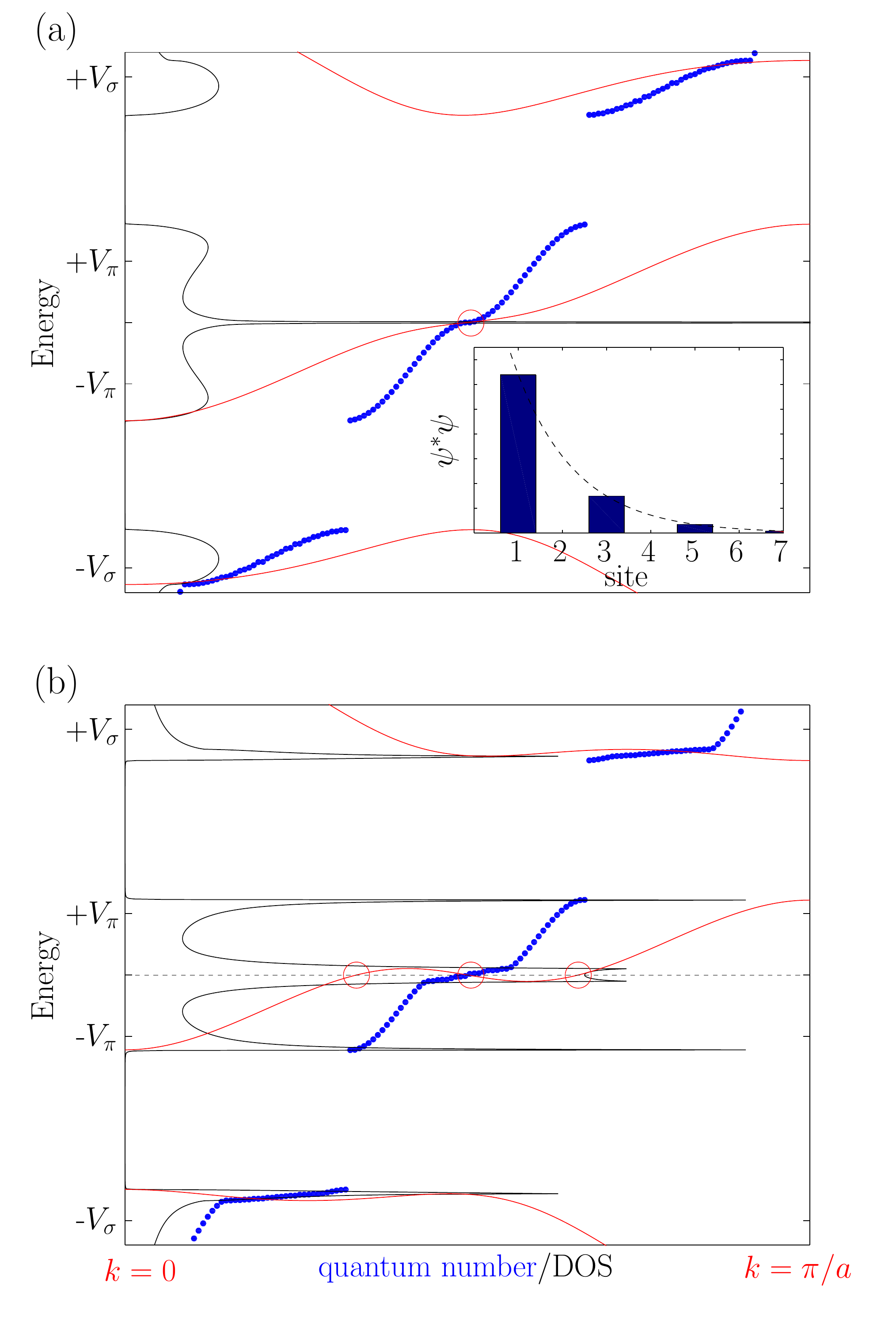}
\caption{Fourfold helix spectrum, $\theta=90^\circ$, with $\eta=4$ using Eq. \ref{eq:t} ($N=3$). Red is bandstructure of the infinite lattice within the irreducible Brillouin zone, black is local DOS of at the end of a semi-infinite lattice, and blue points are the discrete eigenenergies of a 50-site finite lattice. In (a), $\beta=22.5^\circ$ and there is only one band crossing (circled) at zero energy. Assuming time-reversal symmetry (TRS), $\ell=1$. Inset: Probability density for the $\varepsilon=0$ state, showing bipartite structure, and decay controlled by $\min\{\text{eig} (t^{-1}t^\dagger) \}$. In (b), $\beta=33^\circ$ and $0<F<1$ so there are three band crossings ($\ell=3$ assuming TRS) at zero-energy and no bound state exists. \label{fig:spectrum}}
\end{figure}
To confirm our predictions above, we compare the full band structure of an infinite helix obtained by numerical diagonalization of the Bloch Hamiltonian  to the discrete spectrum of a 50-site finite wire. Furthermore, we use numerical solution of the quadratic eigenvalue problem\cite{Dy_PRB79, Brasher_PRB80,Dy_JPC82,Bravi_PRB14} to construct the surface Green's function\cite{Appelbaum_PRB04} $g$ of a semi-infinite wire (with only one endpoint), leading to the density of states (DOS) $\frac{1}{\pi}\Im\{\Tr{g}\}$.

An example spectrum for the fourfold helix is shown in Fig. \ref{fig:spectrum}(a), using parameters $\theta=90^\circ, \beta=22.5^\circ,$ and $\eta=4$. Although it is certainly not evident in the bandstructure (red), and it is concealed within the spectrum of delocalized states of a finite chain (blue), the zero-energy embedded eigenvalue state discussed in Sec. \ref{sec:zero} is clearly visible in the semi-infinite chain DOS (black). The wavefunction probability for this state, obtained from diagonalization of the finite lattice Hamiltonian, is shown in the inset. It has a clear bipartite structure and decay lengthscale as predicted by our analysis of the tight-binding equations.

The spectrum evolves when the parameters are tuned. Importantly, the zero-energy state will disappear when Eq. \ref{eq:zerocondition} is satisfied. The fourfold helix example provides a simple criterion because in this case $F=\frac{(1+\eta)^2}{4\eta}\cos^4\beta$, so that when $\eta=4$, the bound mode will be present only for $\beta<\arccos (2/\sqrt{5})\approx 26.5^{\circ}$. Increasing $\beta$ (as in Fig. \ref{fig:spectrum}(b), where $\beta=33^\circ$) causes a bifurcation of band extrema near zero energy, three band crossings, and a broadening of the bound state corresponding to wavefunction delocalization. The topological zero-energy state has vanished.

\begin{figure}
\includegraphics[scale=0.5,trim={1.7cm 0 0 0},clip]{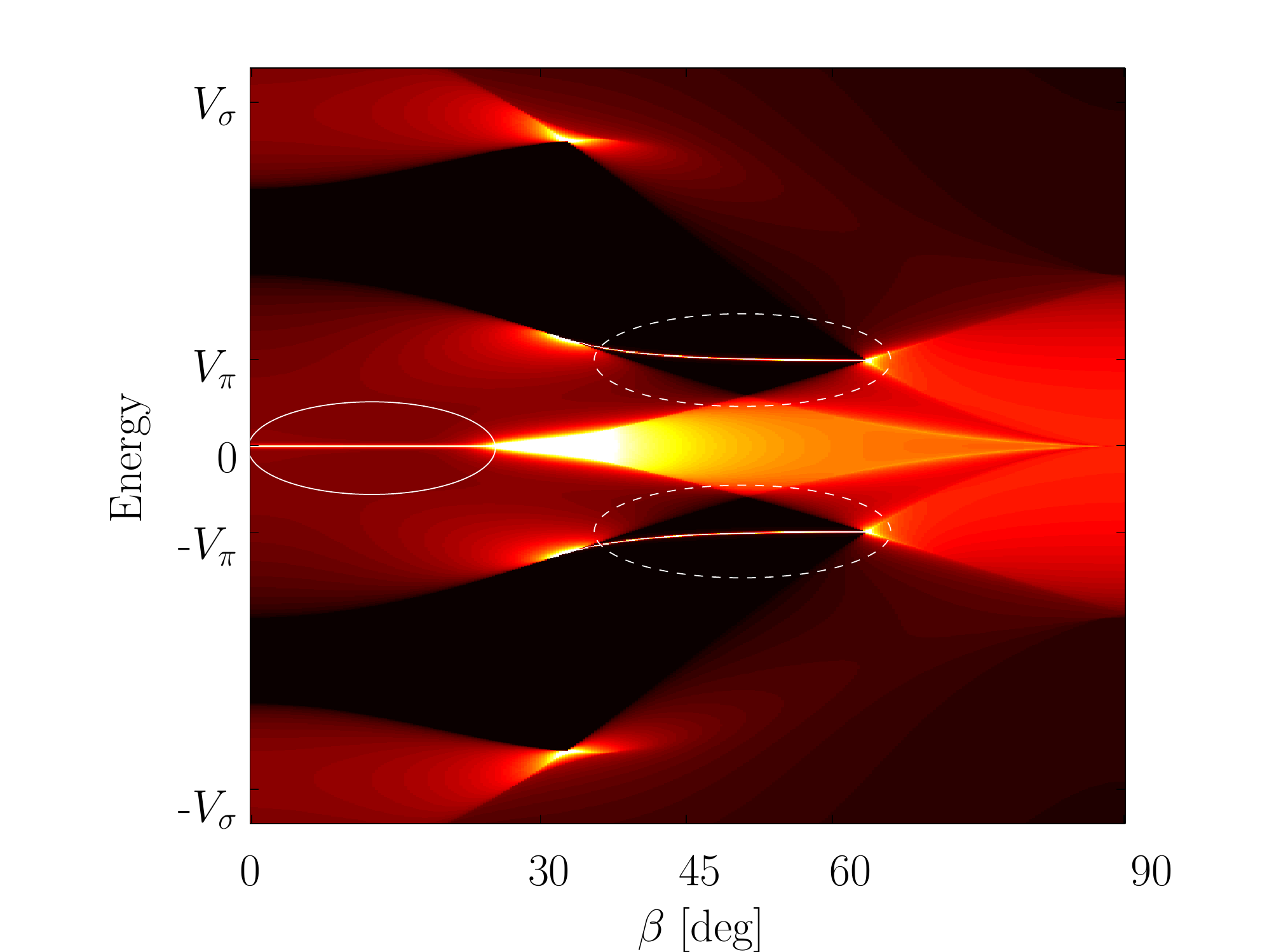}
\caption{Evolution of the spectral density of states for a fourfold helix with $\eta=4$. Black indicates forbidden gap. Note zero-energy mode (indicated by solid ellipse) disappears for $\beta$ above a certain value $\approx 26.5^\circ$; see text. Two gapped bound states are indicated by dashed ellipses. \label{fig:DOSsurf}}
\end{figure}

This evolution can be visualized as a continuous function of $\beta$ by plotting the DOS as in Fig. \ref{fig:DOSsurf}. The zero energy mode, present for small $\beta$ and highlighted by a solid white ellipse, disappears above $\beta\approx 26.5^\circ$ as predicted.

\section{Gapped bound states}
At large values of $\beta$ in Fig.~\ref{fig:DOSsurf}, two new localized modes appear in the spectrum (highlighted by dashed ellipses). These states exist at energy very close to -- but not exactly at -- the coupling parameter $V_\pi$, and do not have the bipartite wavefunction of the topological zero-energy mode. 

In fact, 
all solutions
of the semi-infinite tight-binding Schr\"odinger equation with energy {\em{exactly}} at $\varepsilon=V_\pi$ have {\em{tri}}partite structure, i.e. their wavefunction vanishes at every {\em{third}} site. This property can readily be seen from Eq.~ (\ref{eq:schr}), for $n=1,2$. Combining both to obtain 
\begin{align}
\psi_3=t^{-2}(\varepsilon^2-tt^\dagger)\psi_1,\notag
\end{align}
we notice that, because the hopping matrix $t$ is constructed from unitary rotation matrices applied on the diagonal matrix $D=\text{diag}(-V_\pi,-V_\pi,V_\sigma)$, the transfer operator is singular at $\varepsilon=V_\pi$. As a result, we must have $\psi_3=\mathbf{0}$. An identical argument applies to the next three sites, and so on. We note that this argument also applies to any arbitrary semi-infinite chain with hermitian hopping matrix.

The transfer matrix technique can shed light on this set of gapped states. At nonzero energy, Eq.~(\ref{eq:schr}) can be expressed as 
\begin{align}
\begin{bmatrix}
\psi_{n+1}\\
\psi_n
\end{bmatrix}=
\mathcal{M}(\varepsilon)\begin{bmatrix}
\psi_n\\
\psi_{n-1}
\end{bmatrix},\notag
\end{align}
where the generalized transfer matrix\cite{Dwivedi_PRB16} is given by 
\begin{align}
\mathcal{M}(\varepsilon)=\begin{bmatrix}
\varepsilon t^{-1}&M\\
I_3&0
\end{bmatrix}.\label{eq:genM}
\end{align}
Note that $\mathcal{M}^T \Omega \mathcal{M}=\Omega$, where the skew-symmetric matrix
\begin{align}
\Omega=\begin{bmatrix}
0&-t^T\\
t&0
\end{bmatrix}.\notag
\end{align}
The transfer matrix Eq. \ref{eq:genM} is thus symplectic\cite{Peng_PRB17}.

The eigenvalues of any symplectic matrix come in reciprocal pairs, so that eigenvalue decomposition $M = \mathcal{V}^{-1}\mathcal{D}\mathcal{V}$ has $\mathcal{D} = \text{diag}(z_1,z_2,z_3,z_1^{-1},z_2^{-1},z_3^{-1})$ with $|z_{1,2,3}|> 1$ in the bulk gap. All physical states must be devoid of contributions from their three associated eigenvectors, to prevent divergence of the wavefunction toward the bulk. Considering the boundary condition $[\psi_1,\mathbf{0}]^\dagger$, we have
\begin{align}
\det[\mathcal{V}_{11}(\varepsilon)]=0,
\label{eq:nonzero}
\end{align}
where $\mathcal{V}_{11}(\varepsilon)$ is the upper left $3\times3$ block of $\mathcal{V}$.

\begin{figure}
\includegraphics[scale=0.4]{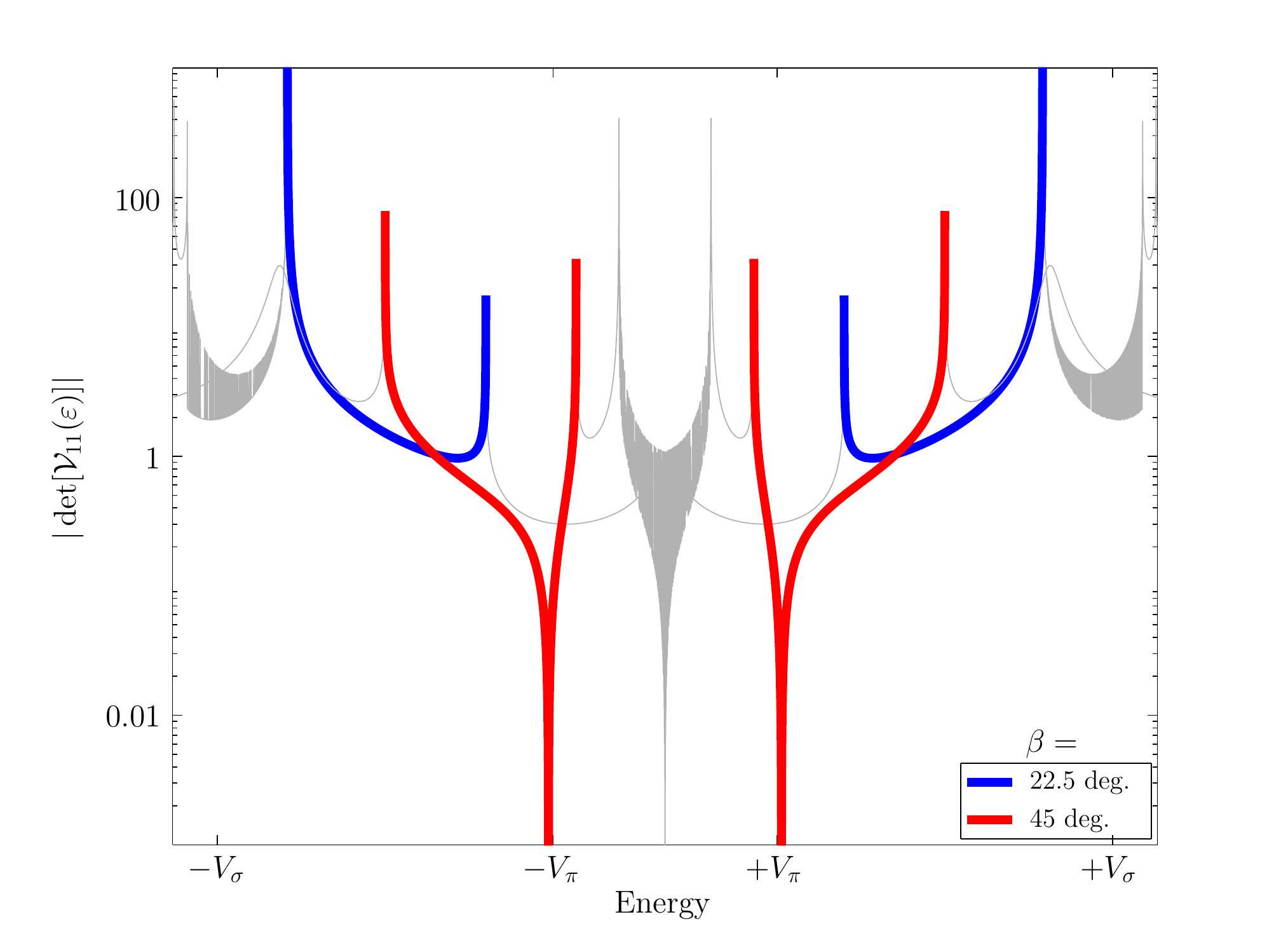}
\caption{ Eq. \ref{eq:nonzero} using $\eta=4$, $\theta=90^\circ$, and two values of $\beta$: $22.5^\circ$ (blue), where no zero exists in the gap, and $45^\circ$ (red), with a zero close to $\varepsilon=\pm V_\pi$ indicating a gapped bound state.  \label{fig:detB}}
\end{figure}

To demonstrate the use of Eq. \ref{eq:nonzero} in providing a criterion for the presence of gapped bound states, we plot $|\det[\mathcal{V}_{11}(\varepsilon)]|$ on a semilog axis in Fig. \ref{fig:detB}, with $\eta=4$ and $\theta=90^\circ$, for $\beta=22.5^\circ$ (where no gapped state can be seen in Fig. \ref{fig:DOSsurf}) and $\beta=45^\circ$ (gapped state present in Fig.~\ref{fig:DOSsurf}). Energy ranges in the bulk gap, where Eq. \ref{eq:nonzero} is valid, are highlighted by the indicated colors. Note that the zero crossing for $\beta=45^\circ$ appears close to $\varepsilon=V_\pi$.

In further work, we will show how a gapped zero-dimensional state of this type disperses into a two-dimensional surface state of bulk trigonal elemental chalcogens. 

\section{Discussion \& Conclusion\label{sec:conc}}

In this paper, we have studied the zero-energy states of chains with helical symmetry. The helical structure of such a chain leads to an effective translational symmetry so that the chain can be mapped to a tight-binding model with one site per unit cell. We find that such chiral models with no on-site term have an interesting topological structure where a semi-infinite chain described by such a Hamiltonian supports topologically protected zero energy states under certain conditions. The number of such states is determined by the number of zero-energy crossings of the bulk bands, which is then identified as the bulk topological invariant describing the topological end modes. 

In the case of an odd number of orbitals per unit cell, these states always co-exist with bulk zero-energy states. While variation in bond strengths and on-site energies close to the endpoint (consistent with broken translational invariance at the boundary) invalidates the Bloch expansion used here, a variation of the transfer matrix approach in Sec. \ref{sec:transfer} can be used to show that the zero-energy bound states remain. Nonzero on-site terms  (such as spin-orbit interaction \cite{Chadi_PRB77}), second-nearest-neighbor coupling, or inclusion of remote $s$ and $d$ orbitals, may perturbatively induce hybridization with bulk states but will merely broaden the DOS peak at zero energy inherited from the topological character of the underlying perfect helix.

We have used the example of $\eta=4$ several times in this paper for illustrative purposes. However, we note  that when $\eta=1$ ($V_\pi=V_\sigma$), Eq. \ref{eq:zerocondition} is always satisfied and there is never a bound state at zero energy for any value of the other parameters $\theta$ and $\beta$. A larger phase space for its existence is enabled by a large (or, mathematically but unphysically, small) value of $\eta$. This makes zero-energy bound states more relevant to helices of elements lower on the periodic table, where $\sigma$ bonding dominates.\cite{Papoian_ACIE00}

We have cast the entire discussion in this paper solely in terms of chains with helical geometry, but our results extend to other lattices with nonzero angular momentum valence orbitals. A simple example is a planar zig-zag chain with 90$^\circ$ bond angles at each site. The $p_x$ and $p_y$ orbitals that lie in the plane and oriented parallel or perpendicular to each bond are, like the four-fold helix, decoupled from each other with alternating nearest-neighbor bonds (from purely $\pi$ to purely $\sigma$). It therefore also has the SSH structure and associated zero-energy bound states.  Although it is clearly not a helix, this planar chain can be considered a pathological case of our model, with $\theta=180^\circ$ and $\beta=45^\circ$, accounting for the $F<0$ violation case shown in Fig. \ref{fig:F}. 

As mentioned earlier, the helix configuration is prevalent in many organic polymers essential for life on earth. This fact tempts us to speculate that the helix geometry serves a distinct biological purpose beyond simply accommodating the stereochemical requirements demanded by the structural dimensions of constituents such as nucleotides in DNA.\cite{Crick_PRS54} 
However, the simplified general model presented in the present work cannot be expected to accurately capture the details of real organic molecules, beyond the mere suggestion that localized bound states are present and relevant to its biological purpose. Extension of our model to capture the specifics of DNA must be performed to test this speculative hypothesis. Whereas the individual atoms (C,N,O,P) have $s,p$-orbital valence electrons, adapting our helix model  beyond the Slater-Koster scheme might use the nucleotide molecular states themselves as a more natural reduced basis. These monomers are not identical, but the nucleobase details play a minor role in electronic structure since the helix is primarily held together by bonds at the phosphate backbone. Thus, the presence of genetic information in the nucleotide sequence can be incorporated by on-site disorder.   

\begin{acknowledgments}
Work by I.A. and P.L. is supported by the Office of Naval Research under contract N000141712994, and the Defense Threat
Reduction Agency under contract HDTRA1-13-1-0013. J.D.S. acknowledges
funding from a Sloan Research Fellowship and NSF DMR-1555135 (CAREER).
\end{acknowledgments}

\hspace{2cm}

\appendix*

\section{On the eigenvalues of \texorpdfstring{$-t^{-1}t^\dagger$}{-inv(t)t'}\label{sec:app}}
Here, $t$ is nonsingular. Let $M = -t^{-1}t^\dagger$, whose determinant is -1. We then have $M^\dagger tM =  [-t(t^{-1})^\dagger]t[-t^{-1}t^\dagger] = t$. Let $\phi$ be the eigenvector of $M$ associated with the eigenvalue $\lambda$: $M\phi = \lambda\phi$. Then, $M^\dagger tM\phi = \lambda M^\dagger t\phi = t\phi$, so that the last equation gives
\begin{align}
M^{\dagger} (t \phi) = \frac{1}{\lambda} (t\phi).
\label{eq:proof_I}
\end{align}
Thus, $\lambda^{-1}$ is the eigenvalue of the operator $M^{\dagger}$ with the associated eigenvector $t\phi$. Since $M$ and $M^{\dagger}$ share the same set of eigenvalues, $\lambda^{-1}$ is also an eigenvalue of $M$.

If $M$ is a $3\times 3$ matrix, simple enumeration of  all three eigenvalues must be $\{z,z^{-1},-1\}$.

\bibliography{helix}

\begin{thebibliography}{24}%
\makeatletter
\providecommand \@ifxundefined [1]{%
 \@ifx{#1\undefined}
}%
\providecommand \@ifnum [1]{%
 \ifnum #1\expandafter \@firstoftwo
 \else \expandafter \@secondoftwo
 \fi
}%
\providecommand \@ifx [1]{%
 \ifx #1\expandafter \@firstoftwo
 \else \expandafter \@secondoftwo
 \fi
}%
\providecommand \natexlab [1]{#1}%
\providecommand \enquote  [1]{``#1''}%
\providecommand \bibnamefont  [1]{#1}%
\providecommand \bibfnamefont [1]{#1}%
\providecommand \citenamefont [1]{#1}%
\providecommand \href@noop [0]{\@secondoftwo}%
\providecommand \href [0]{\begingroup \@sanitize@url \@href}%
\providecommand \@href[1]{\@@startlink{#1}\@@href}%
\providecommand \@@href[1]{\endgroup#1\@@endlink}%
\providecommand \@sanitize@url [0]{\catcode `\\12\catcode `\$12\catcode
  `\&12\catcode `\#12\catcode `\^12\catcode `\_12\catcode `\%12\relax}%
\providecommand \@@startlink[1]{}%
\providecommand \@@endlink[0]{}%
\providecommand \url  [0]{\begingroup\@sanitize@url \@url }%
\providecommand \@url [1]{\endgroup\@href {#1}{\urlprefix }}%
\providecommand \urlprefix  [0]{URL }%
\providecommand \Eprint [0]{\href }%
\providecommand \doibase [0]{http://dx.doi.org/}%
\providecommand \selectlanguage [0]{\@gobble}%
\providecommand \bibinfo  [0]{\@secondoftwo}%
\providecommand \bibfield  [0]{\@secondoftwo}%
\providecommand \translation [1]{[#1]}%
\providecommand \BibitemOpen [0]{}%
\providecommand \bibitemStop [0]{}%
\providecommand \bibitemNoStop [0]{.\EOS\space}%
\providecommand \EOS [0]{\spacefactor3000\relax}%
\providecommand \BibitemShut  [1]{\csname bibitem#1\endcsname}%
\let\auto@bib@innerbib\@empty
\bibitem [{\citenamefont {Glazer}\ and\ \citenamefont
  {Stadnicka}(1986)}]{Glazer_JAC86}%
  \BibitemOpen
  \bibfield  {author} {\bibinfo {author} {\bibfnamefont {A.~M.}\ \bibnamefont
  {Glazer}}\ and\ \bibinfo {author} {\bibfnamefont {K.}~\bibnamefont
  {Stadnicka}},\ }\href {https://doi.org/10.1107/S0021889886089823} {\bibfield
  {journal} {\bibinfo  {journal} {J. Appl. Cryst.}\ }\textbf {\bibinfo {volume}
  {19}},\ \bibinfo {pages} {108} (\bibinfo {year} {1986})}\BibitemShut
  {NoStop}%
\bibitem [{\citenamefont {Tamm}(1932)}]{Tamm_PZSU32}%
  \BibitemOpen
  \bibfield  {author} {\bibinfo {author} {\bibfnamefont {I.}~\bibnamefont
  {Tamm}},\ }\href@noop {} {\bibfield  {journal} {\bibinfo  {journal} {Phys. Z.
  Soviet Union}\ }\textbf {\bibinfo {volume} {1}},\ \bibinfo {pages} {733}
  (\bibinfo {year} {1932})}\BibitemShut {NoStop}%
\bibitem [{\citenamefont {Shockley}(1939)}]{Shockley_PR39}%
  \BibitemOpen
  \bibfield  {author} {\bibinfo {author} {\bibfnamefont {W.}~\bibnamefont
  {Shockley}},\ }\href {\doibase 10.1103/PhysRev.56.317} {\bibfield  {journal}
  {\bibinfo  {journal} {Phys. Rev.}\ }\textbf {\bibinfo {volume} {56}},\
  \bibinfo {pages} {317} (\bibinfo {year} {1939})}\BibitemShut {NoStop}%
\bibitem [{\citenamefont {Su}\ \emph {et~al.}(1979)\citenamefont {Su},
  \citenamefont {Schrieffer},\ and\ \citenamefont {Heeger}}]{Su_PRL79}%
  \BibitemOpen
  \bibfield  {author} {\bibinfo {author} {\bibfnamefont {W.~P.}\ \bibnamefont
  {Su}}, \bibinfo {author} {\bibfnamefont {J.~R.}\ \bibnamefont {Schrieffer}},
  \ and\ \bibinfo {author} {\bibfnamefont {A.~J.}\ \bibnamefont {Heeger}},\
  }\href {\doibase 10.1103/PhysRevLett.42.1698} {\bibfield  {journal} {\bibinfo
   {journal} {Phys. Rev. Lett.}\ }\textbf {\bibinfo {volume} {42}},\ \bibinfo
  {pages} {1698} (\bibinfo {year} {1979})}\BibitemShut {NoStop}%
\bibitem [{\citenamefont {Su}\ \emph {et~al.}(1980)\citenamefont {Su},
  \citenamefont {Schrieffer},\ and\ \citenamefont {Heeger}}]{Su_PRB80}%
  \BibitemOpen
  \bibfield  {author} {\bibinfo {author} {\bibfnamefont {W.~P.}\ \bibnamefont
  {Su}}, \bibinfo {author} {\bibfnamefont {J.~R.}\ \bibnamefont {Schrieffer}},
  \ and\ \bibinfo {author} {\bibfnamefont {A.~J.}\ \bibnamefont {Heeger}},\
  }\href {\doibase 10.1103/PhysRevB.22.2099} {\bibfield  {journal} {\bibinfo
  {journal} {Phys. Rev. B}\ }\textbf {\bibinfo {volume} {22}},\ \bibinfo
  {pages} {2099} (\bibinfo {year} {1980})}\BibitemShut {NoStop}%
\bibitem [{\citenamefont {Heeger}\ \emph {et~al.}(1988)\citenamefont {Heeger},
  \citenamefont {Kivelson}, \citenamefont {Schrieffer},\ and\ \citenamefont
  {Su}}]{Heeger_RMP88}%
  \BibitemOpen
  \bibfield  {author} {\bibinfo {author} {\bibfnamefont {A.~J.}\ \bibnamefont
  {Heeger}}, \bibinfo {author} {\bibfnamefont {S.}~\bibnamefont {Kivelson}},
  \bibinfo {author} {\bibfnamefont {J.~R.}\ \bibnamefont {Schrieffer}}, \ and\
  \bibinfo {author} {\bibfnamefont {W.~P.}\ \bibnamefont {Su}},\ }\href
  {\doibase 10.1103/RevModPhys.60.781} {\bibfield  {journal} {\bibinfo
  {journal} {Rev. Mod. Phys.}\ }\textbf {\bibinfo {volume} {60}},\ \bibinfo
  {pages} {781} (\bibinfo {year} {1988})}\BibitemShut {NoStop}%
\bibitem [{\citenamefont {Hirschle}\ \emph {et~al.}(2001)\citenamefont
  {Hirschle}, \citenamefont {Emmerlinga},\ and\ \citenamefont
  {R\"ohra}}]{Hirschle_ACC01}%
  \BibitemOpen
  \bibfield  {author} {\bibinfo {author} {\bibfnamefont {C.}~\bibnamefont
  {Hirschle}}, \bibinfo {author} {\bibfnamefont {F.}~\bibnamefont
  {Emmerlinga}}, \ and\ \bibinfo {author} {\bibfnamefont {C.}~\bibnamefont
  {R\"ohra}},\ }\href {\doibase 10.1107/S0108270101002451} {\bibfield
  {journal} {\bibinfo  {journal} {Acta Crystallogr. Sect. C}\ }\textbf
  {\bibinfo {volume} {57}},\ \bibinfo {pages} {501} (\bibinfo {year}
  {2001})}\BibitemShut {NoStop}%
\bibitem [{\citenamefont {Doi}\ \emph {et~al.}(1970)\citenamefont {Doi},
  \citenamefont {Nakao},\ and\ \citenamefont {Kamimura}}]{Doi_JPSP70}%
  \BibitemOpen
  \bibfield  {author} {\bibinfo {author} {\bibfnamefont {T.}~\bibnamefont
  {Doi}}, \bibinfo {author} {\bibfnamefont {K.}~\bibnamefont {Nakao}}, \ and\
  \bibinfo {author} {\bibfnamefont {H.}~\bibnamefont {Kamimura}},\ }\href
  {\doibase 10.1143/JPSJ.28.36} {\bibfield  {journal} {\bibinfo  {journal} {J.
  Phys. Soc. Jpn.}\ }\textbf {\bibinfo {volume} {28}},\ \bibinfo {pages} {36}
  (\bibinfo {year} {1970})}\BibitemShut {NoStop}%
\bibitem [{\citenamefont {Joannopoulos}\ \emph {et~al.}(1975)\citenamefont
  {Joannopoulos}, \citenamefont {Schl\"uter},\ and\ \citenamefont
  {Cohen}}]{Joannopoulos_PRB75}%
  \BibitemOpen
  \bibfield  {author} {\bibinfo {author} {\bibfnamefont {J.~D.}\ \bibnamefont
  {Joannopoulos}}, \bibinfo {author} {\bibfnamefont {M.}~\bibnamefont
  {Schl\"uter}}, \ and\ \bibinfo {author} {\bibfnamefont {M.~L.}\ \bibnamefont
  {Cohen}},\ }\href {\doibase 10.1103/PhysRevB.11.2186} {\bibfield  {journal}
  {\bibinfo  {journal} {Phys. Rev. B}\ }\textbf {\bibinfo {volume} {11}},\
  \bibinfo {pages} {2186} (\bibinfo {year} {1975})}\BibitemShut {NoStop}%
\bibitem [{\citenamefont {Hirayama}\ \emph {et~al.}(2015)\citenamefont
  {Hirayama}, \citenamefont {Okugawa}, \citenamefont {Ishibashi}, \citenamefont
  {Murakami},\ and\ \citenamefont {Miyake}}]{Hirayama_PRL15}%
  \BibitemOpen
  \bibfield  {author} {\bibinfo {author} {\bibfnamefont {M.}~\bibnamefont
  {Hirayama}}, \bibinfo {author} {\bibfnamefont {R.}~\bibnamefont {Okugawa}},
  \bibinfo {author} {\bibfnamefont {S.}~\bibnamefont {Ishibashi}}, \bibinfo
  {author} {\bibfnamefont {S.}~\bibnamefont {Murakami}}, \ and\ \bibinfo
  {author} {\bibfnamefont {T.}~\bibnamefont {Miyake}},\ }\href {\doibase
  10.1103/PhysRevLett.114.206401} {\bibfield  {journal} {\bibinfo  {journal}
  {Phys. Rev. Lett.}\ }\textbf {\bibinfo {volume} {114}},\ \bibinfo {pages}
  {206401} (\bibinfo {year} {2015})}\BibitemShut {NoStop}%
\bibitem [{\citenamefont {Yoda}\ \emph {et~al.}(2015)\citenamefont {Yoda},
  \citenamefont {Yokoyama},\ and\ \citenamefont {Murakami}}]{Yoda_SciRep15}%
  \BibitemOpen
  \bibfield  {author} {\bibinfo {author} {\bibfnamefont {T.}~\bibnamefont
  {Yoda}}, \bibinfo {author} {\bibfnamefont {T.}~\bibnamefont {Yokoyama}}, \
  and\ \bibinfo {author} {\bibfnamefont {S.}~\bibnamefont {Murakami}},\ }\href
  {\doibase 10.1038/srep12024} {\bibfield  {journal} {\bibinfo  {journal} {Sci.
  Rep.}\ }\textbf {\bibinfo {volume} {5}},\ \bibinfo {pages} {12024} (\bibinfo
  {year} {2015})}\BibitemShut {NoStop}%
\bibitem [{\citenamefont {Ryu}\ \emph {et~al.}(2010)\citenamefont {Ryu},
  \citenamefont {Schnyder}, \citenamefont {Furusaki},\ and\ \citenamefont
  {Ludwig}}]{Ryu_NJP10}%
  \BibitemOpen
  \bibfield  {author} {\bibinfo {author} {\bibfnamefont {S.}~\bibnamefont
  {Ryu}}, \bibinfo {author} {\bibfnamefont {A.~P.}\ \bibnamefont {Schnyder}},
  \bibinfo {author} {\bibfnamefont {A.}~\bibnamefont {Furusaki}}, \ and\
  \bibinfo {author} {\bibfnamefont {A.~W.~W.}\ \bibnamefont {Ludwig}},\ }\href
  {http://stacks.iop.org/1367-2630/12/i=6/a=065010} {\bibfield  {journal}
  {\bibinfo  {journal} {New J. Phys.}\ }\textbf {\bibinfo {volume} {12}},\
  \bibinfo {pages} {065010} (\bibinfo {year} {2010})}\BibitemShut {NoStop}%
\bibitem [{\citenamefont {von Neumann}\ and\ \citenamefont
  {Wigner}(1929)}]{vonNeumann_PZ29}%
  \BibitemOpen
  \bibfield  {author} {\bibinfo {author} {\bibfnamefont {J.}~\bibnamefont {von
  Neumann}}\ and\ \bibinfo {author} {\bibfnamefont {E.}~\bibnamefont
  {Wigner}},\ }\href@noop {} {\bibfield  {journal} {\bibinfo  {journal} {Phys.
  Z.}\ }\textbf {\bibinfo {volume} {30}},\ \bibinfo {pages} {465} (\bibinfo
  {year} {1929})}\BibitemShut {NoStop}%
\bibitem [{\citenamefont {Hsu}\ \emph {et~al.}(2016)\citenamefont {Hsu},
  \citenamefont {Zhen}, \citenamefont {Stone}, \citenamefont {Joannopoulos},\
  and\ \citenamefont {Solja{\v{c}}i{\'{c}}}}]{Hsu_NRM16}%
  \BibitemOpen
  \bibfield  {author} {\bibinfo {author} {\bibfnamefont {C.~W.}\ \bibnamefont
  {Hsu}}, \bibinfo {author} {\bibfnamefont {B.}~\bibnamefont {Zhen}}, \bibinfo
  {author} {\bibfnamefont {A.~D.}\ \bibnamefont {Stone}}, \bibinfo {author}
  {\bibfnamefont {J.~D.}\ \bibnamefont {Joannopoulos}}, \ and\ \bibinfo
  {author} {\bibfnamefont {M.}~\bibnamefont {Solja{\v{c}}i{\'{c}}}},\ }\href
  {\doibase 10.1038/natrevmats.2016.48} {\bibfield  {journal} {\bibinfo
  {journal} {Nat. Rev. Mater.}\ }\textbf {\bibinfo {volume} {1}},\ \bibinfo
  {pages} {16048} (\bibinfo {year} {2016})}\BibitemShut {NoStop}%
\bibitem [{\citenamefont {Dy}\ \emph {et~al.}(1979)\citenamefont {Dy},
  \citenamefont {Wu},\ and\ \citenamefont {Spratlin}}]{Dy_PRB79}%
  \BibitemOpen
  \bibfield  {author} {\bibinfo {author} {\bibfnamefont {K.~S.}\ \bibnamefont
  {Dy}}, \bibinfo {author} {\bibfnamefont {S.-Y.}\ \bibnamefont {Wu}}, \ and\
  \bibinfo {author} {\bibfnamefont {T.}~\bibnamefont {Spratlin}},\ }\href
  {\doibase 10.1103/PhysRevB.20.4237} {\bibfield  {journal} {\bibinfo
  {journal} {Phys. Rev. B}\ }\textbf {\bibinfo {volume} {20}},\ \bibinfo
  {pages} {4237} (\bibinfo {year} {1979})}\BibitemShut {NoStop}%
\bibitem [{\citenamefont {Brasher}\ and\ \citenamefont
  {Dy}(1980)}]{Brasher_PRB80}%
  \BibitemOpen
  \bibfield  {author} {\bibinfo {author} {\bibfnamefont {J.~D.}\ \bibnamefont
  {Brasher}}\ and\ \bibinfo {author} {\bibfnamefont {K.~S.}\ \bibnamefont
  {Dy}},\ }\href {\doibase 10.1103/PhysRevB.22.4868} {\bibfield  {journal}
  {\bibinfo  {journal} {Phys. Rev. B}\ }\textbf {\bibinfo {volume} {22}},\
  \bibinfo {pages} {4868} (\bibinfo {year} {1980})}\BibitemShut {NoStop}%
\bibitem [{\citenamefont {Dy}\ and\ \citenamefont {Brasher}(1982)}]{Dy_JPC82}%
  \BibitemOpen
  \bibfield  {author} {\bibinfo {author} {\bibfnamefont {K.~S.}\ \bibnamefont
  {Dy}}\ and\ \bibinfo {author} {\bibfnamefont {J.~D.}\ \bibnamefont
  {Brasher}},\ }\href {http://stacks.iop.org/0022-3719/15/i=3/a=027} {\bibfield
   {journal} {\bibinfo  {journal} {J. Phys. C}\ }\textbf {\bibinfo {volume}
  {15}},\ \bibinfo {pages} {633} (\bibinfo {year} {1982})}\BibitemShut
  {NoStop}%
\bibitem [{\citenamefont {Bravi}\ \emph {et~al.}(2014)\citenamefont {Bravi},
  \citenamefont {Farchioni}, \citenamefont {Grosso},\ and\ \citenamefont
  {Pastori~Parravicini}}]{Bravi_PRB14}%
  \BibitemOpen
  \bibfield  {author} {\bibinfo {author} {\bibfnamefont {M.}~\bibnamefont
  {Bravi}}, \bibinfo {author} {\bibfnamefont {R.}~\bibnamefont {Farchioni}},
  \bibinfo {author} {\bibfnamefont {G.}~\bibnamefont {Grosso}}, \ and\ \bibinfo
  {author} {\bibfnamefont {G.}~\bibnamefont {Pastori~Parravicini}},\ }\href
  {\doibase 10.1103/PhysRevB.90.155445} {\bibfield  {journal} {\bibinfo
  {journal} {Phys. Rev. B}\ }\textbf {\bibinfo {volume} {90}},\ \bibinfo
  {pages} {155445} (\bibinfo {year} {2014})}\BibitemShut {NoStop}%
\bibitem [{\citenamefont {Appelbaum}\ \emph {et~al.}(2004)\citenamefont
  {Appelbaum}, \citenamefont {Wang}, \citenamefont {Joannopoulos},\ and\
  \citenamefont {Narayanamurti}}]{Appelbaum_PRB04}%
  \BibitemOpen
  \bibfield  {author} {\bibinfo {author} {\bibfnamefont {I.}~\bibnamefont
  {Appelbaum}}, \bibinfo {author} {\bibfnamefont {T.}~\bibnamefont {Wang}},
  \bibinfo {author} {\bibfnamefont {J.~D.}\ \bibnamefont {Joannopoulos}}, \
  and\ \bibinfo {author} {\bibfnamefont {V.}~\bibnamefont {Narayanamurti}},\
  }\href {\doibase 10.1103/PhysRevB.69.165301} {\bibfield  {journal} {\bibinfo
  {journal} {Phys. Rev. B}\ }\textbf {\bibinfo {volume} {69}},\ \bibinfo
  {pages} {165301} (\bibinfo {year} {2004})}\BibitemShut {NoStop}%
\bibitem [{\citenamefont {Dwivedi}\ and\ \citenamefont
  {Chua}(2016)}]{Dwivedi_PRB16}%
  \BibitemOpen
  \bibfield  {author} {\bibinfo {author} {\bibfnamefont {V.}~\bibnamefont
  {Dwivedi}}\ and\ \bibinfo {author} {\bibfnamefont {V.}~\bibnamefont {Chua}},\
  }\href {\doibase 10.1103/PhysRevB.93.134304} {\bibfield  {journal} {\bibinfo
  {journal} {Phys. Rev. B}\ }\textbf {\bibinfo {volume} {93}},\ \bibinfo
  {pages} {134304} (\bibinfo {year} {2016})}\BibitemShut {NoStop}%
\bibitem [{\citenamefont {Peng}\ \emph {et~al.}(2017)\citenamefont {Peng},
  \citenamefont {Bao},\ and\ \citenamefont {von Oppen}}]{Peng_PRB17}%
  \BibitemOpen
  \bibfield  {author} {\bibinfo {author} {\bibfnamefont {Y.}~\bibnamefont
  {Peng}}, \bibinfo {author} {\bibfnamefont {Y.}~\bibnamefont {Bao}}, \ and\
  \bibinfo {author} {\bibfnamefont {F.}~\bibnamefont {von Oppen}},\ }\href
  {\doibase 10.1103/PhysRevB.95.235143} {\bibfield  {journal} {\bibinfo
  {journal} {Phys. Rev. B}\ }\textbf {\bibinfo {volume} {95}},\ \bibinfo
  {pages} {235143} (\bibinfo {year} {2017})}\BibitemShut {NoStop}%
\bibitem [{\citenamefont {Chadi}(1977)}]{Chadi_PRB77}%
  \BibitemOpen
  \bibfield  {author} {\bibinfo {author} {\bibfnamefont {D.~J.}\ \bibnamefont
  {Chadi}},\ }\href {\doibase 10.1103/PhysRevB.16.790} {\bibfield  {journal}
  {\bibinfo  {journal} {Phys. Rev. B}\ }\textbf {\bibinfo {volume} {16}},\
  \bibinfo {pages} {790} (\bibinfo {year} {1977})}\BibitemShut {NoStop}%
\bibitem [{\citenamefont {A.~Papoian}\ and\ \citenamefont
  {Hoffmann}(2000)}]{Papoian_ACIE00}%
  \BibitemOpen
  \bibfield  {author} {\bibinfo {author} {\bibfnamefont {G.}~\bibnamefont
  {A.~Papoian}}\ and\ \bibinfo {author} {\bibfnamefont {R.}~\bibnamefont
  {Hoffmann}},\ }\href {\doibase
  10.1002/1521-3773(20000717)39:14<2408::AID-ANIE2408>3.0.CO;2-U} {\bibfield
  {journal} {\bibinfo  {journal} {Angew. Chem. Int. Ed.}\ }\textbf {\bibinfo
  {volume} {39}},\ \bibinfo {pages} {2408} (\bibinfo {year}
  {2000})}\BibitemShut {NoStop}%
\bibitem [{\citenamefont {Crick}\ and\ \citenamefont
  {Watson}(1954)}]{Crick_PRS54}%
  \BibitemOpen
  \bibfield  {author} {\bibinfo {author} {\bibfnamefont {F.}~\bibnamefont
  {Crick}}\ and\ \bibinfo {author} {\bibfnamefont {J.}~\bibnamefont {Watson}},\
  }\href@noop {} {\bibfield  {journal} {\bibinfo  {journal} {Proc. Royal Soc.
  London}\ }\textbf {\bibinfo {volume} {223}},\ \bibinfo {pages} {80} (\bibinfo
  {year} {1954})}\BibitemShut {NoStop}%
\end{thebibliography}%

\end{document}